# Structures and Dynamics of Lone Schur Flows with Vorticity but no Swirls

Jian-Zhou Zhu (朱建州)@sccfis.org

**Abstract**

We study the dynamics and indications of the flows with all the eigenvalues of the velocity gradients being real, thus 'lone', *i.e.*, without forming the complex conjugate pairs associated to the swirls. A generic prototype is the 'lone Schur flow (LSF)' whose velocity gradient tensor is uniformly of Schur form but free of complex eigenvalues. A (partial) integral-differential equation governing such LSF is established, and a semi-analytical algorithm is accordingly designed for computation. Simulated evolutions of example LSFs in 2- and 3-spaces show rich dynamics and vortical structures, but no obvious swirls (nor even the homoclinic loops in whatever distorted forms) could be found. We discovered the flux loop scenario and the anisotropic analogy of the incompressible turbulence at or close to the critical dimension $D_c = 4/3$ decimated from 2-space.

## Text

A simple planar Couette flow is vortical but with no swirls, and a point vortex flow or, more generally, flows with singular curves present swirls (as the Bose-Einstein condensates modeled by the Gross-Pitaevskii equation [1] with phase defects in 3-space): such are text-book examples but not the generic viscous flows, in the latter of which the genesis and control of swirls are important and difficult problems. For example, swirling motions of fluids, ranging from (ultra-)cold atoms, to classical flows in our daily lives or industries, and to controlled fusion and asstrophysical plasmas, are ubiquitous and intriguing phenomena. The genesis and sustanance of such 'vortexes' as the well-known Great Red Spot or the recently photographed octagonal and pentagonal cyclone clusters encircling, respectively, the northern and southern poles of Jupiter [2], remains mysterious in some sense. In particular, we may ask the opposite question: can the flow be free of swirls relative to a (fast) rotating frame?

The velocity gradient tensor (VGT) plays a central role in solvable and physical models of turbulence fundamentals and phenomenologies (*e.g.*, [3, 4, 5]). While on the very basic side, people have been puzzled by the issue of the identification of vortex and have resorted to the phase portraits of dynamical systems defined by VGT (e.g., Ref. [6] in which the authors express the opinion that "it is unlikely that any definition of a vortex will win universal acceptance.") And, it appears to us that the pure kinematic identification without specifying the definite hydrodynamics is fundamentally incomplete, which underlies the real difficulty of the problem. For instance, complex eigenvalues of the VGT are associated to foci or centers, thus spirals or rotations, and have been considered by various authors in identifying swirling vortexes (*e.g.*, [6]), but with unsatisfying features [7] (*c.f.*, footnote 6 of a parallel communication [8]). However, in our opinion, neither the proposal nor the objection makes complete sense without referring to the specific dynamics.

Since a matrix can always be transformed into the real Schur form [9], a 'real Schur flow (RSF)' uniformly of such VGT matrix could play a role in fundamental hydrodynamics and turbulence resembling that of special relative in general relativity, in the author's belief [10, 11, 12]: in somewhat more mathematical language, the RSF solutions may form a kind of 'generating subspace' of NSF, with the generating rule, serving as the physical 'equivalence principle', thus other possible physics separately embedded in the (real) Schur transformations. Now, the dynamical framework and tool are established to directly attack some of the problems, and the first key results are reported here: the basic idea is to analyze the "'lone' Schur flow (LSF)" whose VGT is of the Schur form ('real' or not) uniformly over space and time, but with the VGT eigenvalues being all real, thus 'lone', *i.e.*, without forming a complex conjugate pair.

For the velocity vector $\boldsymbol{u} := \{u_1, u_2, u_3\}$, with the index ',$i$' $\leftrightarrow$ '$\partial_{x_i}$', the matrix representation of $\nabla \boldsymbol{u}$ in the 'Schur frame' (with appropriate coordinate transformation when necessary [9]),

$$\begin{pmatrix} \lambda_1 & \gamma_1 & \gamma_2 \\ 0 & \lambda_2 & \gamma_3 \\ 0 & 0 & \lambda_3 \end{pmatrix} \text{ or } \begin{pmatrix} u_{1,1} & u_{2,1} & u_{3,1} \\ \cancel{u_{1,2}} & u_{2,2} & u_{3,2} \\ \cancel{u_{1,3}} & \cancel{u_{2,3}} & u_{3,3} \end{pmatrix}. \qquad (0L,0R)$$

Associated to the RSF, $u_{1,2}$ in (0R) does not necessarily vanish (while $u_{1,3} = u_{2,3} \equiv 0$), with a conjugate 'pair' of complex eigenvalues $\lambda_1 = \lambda_2^*$ besides the real one $\lambda_3$ in (0L) where $\gamma_1$, $\gamma_2$ and $\gamma_3$ are real, but when all eigenvalues are real, $\lambda_1 = u_{1,1}$, $\lambda_2 = u_{2,2}$, $\lambda_3 = u_{3,3}$ and $u_{1,2} = 0$ as indicated by the slash for LSF. It is trivial to reduce to the 2-space LSF by considering only the upper-left $2 \times 2$ (sub)matrix or letting $u_{3,3} = \lambda_3 \equiv 0$.

Other issues about the RSF with VGT allowing complex eigenvalues are discussed in Ref. [8] where the 'flux loop' scenario as in two-dimensional (2D) turbulence with stratification [13] or compressibility [14] are found to be responsible for sustaining nonuniversal large-scale vortex. Related 2D arguments can be traced back to Onsager [15] and Kraichnan [16] (K67), among others, but a systematic understanding is still not available. The current LSF results, as those of RSF [8], show that the largest scales of the system driven at small scales depend on the forcing/acceleration schemes. And, the dynamical behavior of LSF is reminiscent of the critical dimension, argued to be $D_c = 4/3$ where the generalized K67 absolute-equilibrium and Kolmogorov 1941 $k^{-5/3}$-cascade spectra meet [17], and of the decimation [18]: LSF will also be shown to be of some 'critical' behaviors, anisotropic though.

Incidentally, concerning the dynamics of VGT and the connections with turbulence physics, misinterpretation of the "processes relate to the turbulent energy cascade" can easily arise and be dynamically misleading, as emphasized by Carbone and Bragg [19] who avoid the claim of 'physical mechanism' which would 'require an explanation for why' the associated properties are such and such. The lesson applies to flows in both 2- and 3-spaces. Thus, though it is presumably informative (see also Ref. [20] and references therein) to compare such VGT relevant properties of RSF, LSF and NSF, we resist rushing immediately into such measurements and making quick claims before careful systematic analysis which is being initiated. Also promising are the comparative studies in such systems the combustion, dynamo and aeroacoustics, as partly indicated in Ref. [11]. Vieillefosse [21] found the finite-time singularity of the restricted Euler equation model proposed there by working in the principle frame of the shear rate matrix where some of Cantwell's [22] exact results are presented as well (see also Carbone and Bragg [23] for the local reduction of dimensional resuction of the pressure Hessian, and references therein for others), and some possibility of similar fashion was speculated in footnote 6 of Ref. [10], deserving further consideration.

Although some of the results are applicable also for more general gases as those of RSF [12], here we restrict to the barotropic case for simplicity; and, for the same reason, appropriate viscosity and acceleration won't affect the following analysis, thus are neglected for the time being. That is, we start from the Euler equation

$$\partial_t \rho + \nabla \cdot (\rho \boldsymbol{u}) = 0, \tag{1}$$

$$\partial_t \boldsymbol{u} + \boldsymbol{u} \cdot \nabla \boldsymbol{u} = -\rho^{-1} \nabla p. \tag{2}$$

For the barotropic case, we can introduce the specific enthalpy $\Pi$ (up to a constant), $\nabla \Pi := (\nabla p)/\rho$, and the isothermal (constant-temperature) relation $p = c^2 \rho$ results in $\nabla \Pi = c^2 \nabla \ln \rho$, or, up to irrelevant constant, simply $\Pi = c^2 \ln \rho$, where $c$ is the sound speed.

Now, consider the LSF with

$$u_{1,2} = u_{1,3} = u_{2,3} \equiv 0. \tag{3}$$

We have one-dimensonal $u_1(x_1)$, two-dimensional $u_2(x_1, x_2)$ and three-dimensional $u_3(x_1, x_2, x_3)$, a '1D2D3D' flow. Fourier representation can be applied for periodic solutions, and an example field $\boldsymbol{u}(t_0)$ is

$$\begin{pmatrix} a_1 \sin(\,^1k_1 x_1) \\ a_2 \cos(\,^2k_1 x_1)\sin(\,^2k_2 x_2) \\ a_3 \cos(\,^3k_1 x_1)\cos(\,^3k_2 x_2)\sin(\,^3k_3 x_3) \end{pmatrix}, \tag{4}$$

or the (random) superpositions of such ansatz or others, which can be used as the initial and/or the acceleration fields in the simulations and where $a_\bullet$ and $^\bullet k_\bullet$ are spatially uniform real coefficients. Fig. 1 presents an example initial field, with $a_\bullet = \,^\bullet k_\bullet = 1$ in Eq. (4), and the decay from it (starting from time $t_0$, according to the LSF dynamics with viscosity to be established below) at the time when small scales are well excited. We see well-structured vorticity ($\nabla \times \boldsymbol{u}$) isosurface patterns, but, as checked also from various other orientation views (Supplementary Materials), no swirling structures, large '(anti)cyclones' or small 'eddies', can be observed from the velocity streamlines. The decay from the same field of RSF and NSF, in both 2- and 3-spaces, however obviously generate swirls (not shown), as should be expected. The absence of swirls in LSFs will be further examined dynamically.

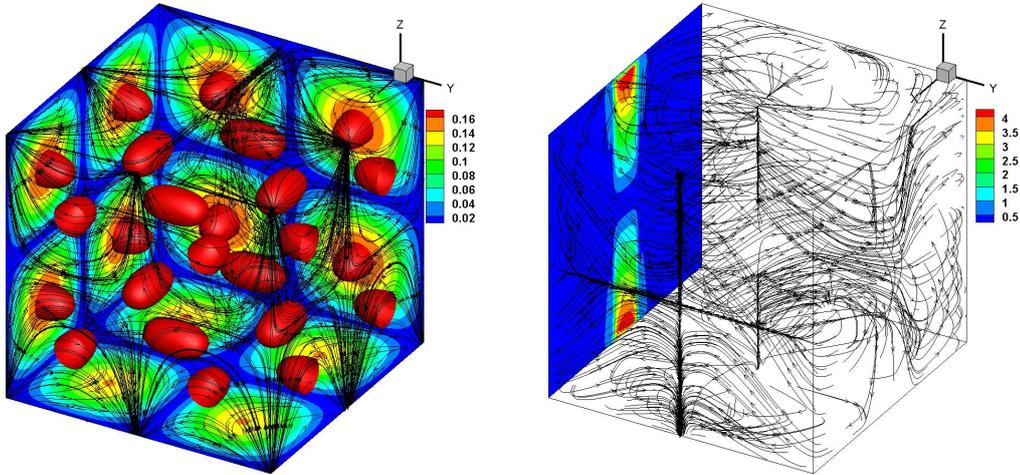

Figure 1: Isosurfaces of vorticity amplitudes and velocity streamtraces of LSF given by Eqs. (4) as the initial field (left) and decaying to a state with developed multi-scale excitations (right).

The establishment of the governing equations and the computational algorithm for LSF can be concisely described as follows, while that for those of RSF in similar fashion can be found in Refs. [12] and [8] which offer more details for comparison.

First, it is not hard to see that Eq. (3) requires

$$\Pi = \mathscr{P}_3(x_3,t) + \mathscr{P}_2(x_2,t) + \mathscr{P}_1(x_1,t). \tag{5}$$

We then consider the LSF in a box of dimension $L_z \times L_2 \times L_3$, cyclic in each direction. Introducing

$$\langle \bullet \rangle_{123} := \frac{\int\int\int \bullet d^3\boldsymbol{x}}{L_1 L_2 L_3} \text{ and } \langle \bullet \rangle_{ij} := \frac{\int\int \bullet dx_i dx_j}{L_i L_j}, \tag{6}$$

with $i$, $j = 1$, 2 and 3, we have

$$\Pi = \langle \Pi \rangle_{23} + \langle \Pi \rangle_{13} + \langle \Pi \rangle_{12} - 2\langle \Pi \rangle_{123}. \tag{7}$$

Taking $\varrho = \boldsymbol{u} \cdot \nabla \ln \rho + \nabla \cdot \boldsymbol{u}$ we obtain further from Eqs. (1 and 2) for the isothermal flow, or even the nonbarotropic ideal gas with $p = \rho RT$ with accompanying structures of $T$, as for RSF [12], the (partial) integral differential equation

$$\partial_t \ln \rho = 2\langle \varrho \rangle_{123} - \langle \varrho \rangle_{23} - \langle \varrho \rangle_{13} - \langle \varrho \rangle_{12}. \tag{8a}$$

$$\partial_t u_1 + u_1 u_{1,1} = -c^2 (\ln \rho)_{,1}, \tag{8b}$$

$$\partial_t u_2 + u_1 u_{2,1} + u_2 u_{2,2} = -c^2 (\ln \rho)_{,2}, \tag{8c}$$

$$\partial_t u_3 + \boldsymbol{u} \cdot \nabla u_3 = -c^2 (\ln \rho)_{,3}. \tag{8d}$$

The above derivation works also for the case with appropriate external acceleration and internal viscosity, and the results may be used in designing the algorithm for computation: the latter is assured by the observation that, given the initial LSF, the above system evolves still to an LSF at a later time.

Our computational algorithm for LSF, integrating directly Eqs. (8a,8b,8c and 8d) with the replacement

$$u_1 = \langle u_1 \rangle_{23} \text{ and } u_2 = \langle u_2 \rangle_3, \tag{9}$$

is *semi-analytical* in the sense that the relations (5) [or (7)] and (3) are satisfied "precisely" (up to the computer round off errors, independent of the numerical errors from whatever methods). And, high-order finite difference schemes are used for spatial discretization (with 128 homogeneous grids in each direction) and time marching was made with third order Runge-Kutta method (see Ref. [8] for such same details). All variables are nondimensionalized with appropriate normalization and the Reynolds number for all cases to be presented is set to be $Re = 350$. Scuh a Reynolds number is not high but the acceleration are added with randomness (in time) in the coefficients, reaching a quasi-steady state, thus the numerical results with developed multi-scale excitations may well be regarded as 'turbulence' in the common sense (though we do not have a consensus on the definition yet). A combination of logarithmic density and conservative-form methods, due to, respectively, the form of Eq. (8a) and the necessity of capturing shocks for cases of high turbulent Mach number ($Ma = 1$): such details are common to RSF and is elaborated in Ref. [8].

To prepare for the discussions of the computational results, we should address the structure of the dynamical equations. First of all, on the one hand, the horizontal vortical mode of LSF is 2D (with the presure terms eliminated by curling the $\boldsymbol{u}_h = \{u_1, u_2\}$ equations to obtain the vertical vorticity equation), thus in principle such a system can support a 'flux loop' as discovered in 2D compressible flows [14], resembling that in 2D stratified turbulence [13], with additional 3D transfer channels now, just as the general RSF studied in Refs. [12, 8]; on the other hand, the vertical vorticity equation is 'lamed' now, with $u_{1,2} = 0$ and with no contribution from Eq. (8b), thus the intrinsic mechanism driving the inverse transfer of 2D solenoidal excitations and that for the flux loop may not be as efficient as that of general RSFs; also, the coupling of the $\boldsymbol{u}_h$ and $u_3$ dynamics through the pressure/density gradients is presumably weaker, due to less quadratic interactions in Eq. (8a): in $\langle \varrho \rangle_1$ we now won't have the quadratic interacton $\langle u_1 (\ln \rho)_{,1} \rangle_1$ which however survives in a general RSF.

Numerical tests of the above remarks were performed with both decaying and driven cases and more analytical arguments on the statistical dynamic 'mechanism' follow. Just as the analysis for the incompressible RSF in 3-space [10], it is shown that solutions strictly periodic in $x_2$ is not allowed for incompressible LSF in 2-space with uniformly vanishing $u_{1,2}$, but it is helpful to still use finite Fourier modes for approximation and formally adapt Kraichnan's [16] (K67) argument with the statistical absolute equilibria for the inverse energy cascade of incompressible 2D flows. We consider the constant of motion, $C = c_E E + c_W W$ where $c_E$ and $c_W$ are the 'temperature' parameters and where the kinetic energy and enstrophy are, respectively,

$$E = \frac{1}{2} \sum_{\boldsymbol{k}} |\hat{u}_1(\boldsymbol{k})|^2 \delta_{0,k_2} + |\hat{u}_2(\boldsymbol{k})|^2 \tag{10}$$

$$W = \frac{1}{2} \sum_{\boldsymbol{k}} (k_1^2 + k_2^2)[|\hat{u}_1(\boldsymbol{k})|^2 \delta_{0,k_2} + |\hat{u}_2(\boldsymbol{k})|^2]. \tag{11}$$

The Kronecker delta $\delta_{0,k_2}$ then also characterizes the anisotropic canonical ensemble for the presumed absolute equilibrium spectra [16]

$$U_1(k) := \langle |\hat{u}_1|^2 \rangle / 2 = \delta_{0,k_2}/(c_E + c_W k^2), \tag{12}$$

$$U_2(k) := \langle |\hat{u}_2|^2 \rangle / 2 = 1/(c_E + c_W k^2). \tag{13}$$

Such anisotropic statistical absolute equilibria (with 'negative temperature' parameter $c_E$) do still indicate the possibility of sustaining large-scale concentration of the kinetic energy. Now, the fact that only $U_2$ concentrates at small $k$ for $k_2 \ne 0$ also indicates weaker effects of condensation. We then expect that the mode interactions leading to the inverse transfer of LSF are much weaker than the complete NSF.

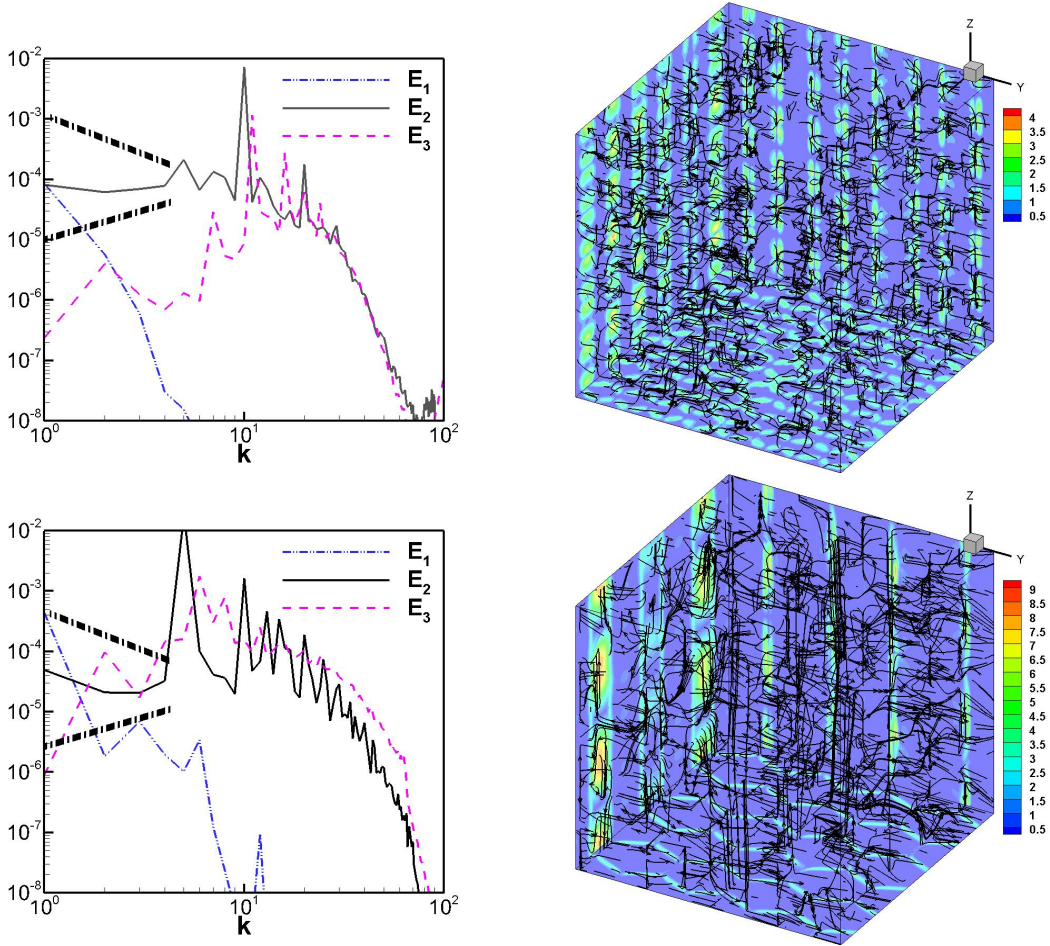

Figure 2: Left: Power spectra of velocity components, accelerated at $k \approx 10$ ($Ma = 0.1$: upper row) and $k \approx 5$ ($Ma = 1$: lower row), with the thick dash-dot lines of, respectively, $k^{-5/3}$ (K67) and $k$ (equipartition in 2-space) spectra indicating that $E_2$ appear in between; right: Isosurfaces of vorticity amplitudes and velocity streamtraces corresponding to the left-panel spectra.

The upper-left panel of Fig. 2 presents the power spectra $E_1$, $E_2$ and $E_3$ of, respectively, $u_1$, $u_2$ and $u_3$ of a chosen snapshot in the (statistical) steady state with $E = E_1 + E_2 + E_3$ close to the time average: the accel-eration obeys Eq. (4), with $k \approx 10$ ($^1k_1 = 10$, $^2k_1 = 6$, $^2k_2 = 8$, $^3k_1 = 6$, $^3k_2 = 6$ and $^3k_3 = 6$) and with random (in time) $a_1$, $a_2$ and $a_3$ being of identical independent distribution uniformly over $(0.5, 1.5]$, and the Mach number is 0.1. As for the general RSF [8], the viscosity model is chosen to be $Re^{-1}\nabla^2 \bm{u}$. The initial field also obey Eq. (4) with $^{\bullet}k_{\bullet} = 1$ and now $a_1 = a_2 = a_3 = 10^{-1}$. It is seen that the excitations at the lowest modes can be sustained, with the establishment of the other excitations in between and resembling the dual-cascade scenario of 2D incompressible turbulence, as also found in compressible and stratified 2D turbulence with flux loops [13, 14]. The conclusion is further supported by the $Ma = 1.0$ case [$k \approx 5$ ($^1k_1 = 5$, $^2k_1 = 3$, $^2k_2 = 4$, $^3k_1 = 3$, $^3k_2 = 3$ and $^3k_3 = 3$) and the rest of the setup being the same] shown in the lower row, with quantitative differences though. The strong piling up of $E_2$ at the driven wavenumbers for both cases is by itself a clear indication that the injected $E_2$ is transferred out inefficiently in either direction. [$E_1$ droping down quickly with increasing $k$ can also be understood by the fact that the fraction of mode numbers with $k_2 = k_3 = 0$ for $u_1$ decrease, roughly as $k^{-1}$ in 2-space and as $k^{-2}$ in 3-space of $\bm{k}$, with the fluctuations of $u_1$ easily spread out of the $k_1$-axis as $u_2$ and $u_3$ fluctuations, except for those condensate at large scales. The $E_3$ behavior is not surprising, since there is not large-scale sustaining mechanism for it. And, many other details can be understood better by checking the power spectra of pressure gradients and the parallel and perperdicular components, among others as the more comprehensive analysis for RSF in Ref. [8], which, together with other issues such as the helicity fastening effect [25], is far beyond the scope of this note and will be communicated elsewhere.] Actually, we have also tried various acceleration schemes with vanishing or 'very small' (purely emperical so far, but intuitively understood to be relative to the strength of the accelerations) initial fields at large scales where we never saw the establishment of velocity power spectra to a level comparable to those around the driven wave numbers (respectively 10 and 5 for the two cases) in 3-space (in general more than 4 orders of magnitude lower in our setups); but in 2-space, we generally did see the establishment of largest-scale excitations to a level even higher than those around the driven wavenumver. Since the dynamics (thus the spectra) are highly anisotropic with each component of velocity being also distinct, and, because of the largest-scale nonuniversality with respect to the acceleration scheme as in RSF [8], many other interesting issues are left to be elaborated elsewhere. But, the clear and important message for this note is the following: the LSF inverse transfer mechanism is weaker than RSF in 3-space to the 'critical' level in some sense analogous to that of fractal dimension $D_c = 4/3$ decimated from 2-space isotropically [18]).

Also importantly, neither any swirl pattern nor closed velocity streamlines (not even in whatever distorted way as the homoclinic loops, say) are observable in the right panels of Fig. 2 of the fields, which is also confirmed with other orientation views (Supplementary Materials). From the knowledge of dynamical systems, the absence of local swirls or closed streamlines is obvious for lack of foci or centers in LSF, but, for large-scale or global behavior in 3-space, the above observation appears nontrivial, constituting the other major result besides the above claimed 'critical' behavior concerning the inverse transfer to form the flux loop. In accordance with the question at the end of the first paragraph, since a compressible flow formally converges to an RSF in the fast rotating limit [11], such particular properties of LSF (which can be regarded as a particular RSF) is of direct fundamental physical relevance (on cyclone genesis, say) and can be of applications in flow control (in aeroacoustics, say).

Finally, let us emphasize that LSF is by definition anisotropic, even in the $x_1$-$x_2$ plane. The quadratic interaction terms in Eq. (8c) for $u_2$ shows that the dynamics of $u_2$ is also intrinsically anisotropic. Thus, also due to the weak fluxes (presumably vanishing for the 'critical' system), the quantification of various spectral transfers for the large-scale flux loop scenario is difficult. Resorting to some closure theory, as in Frisch et al. [18], may help, but by itself is still highly nontrivial due to the anisotropy. More detailed analysis of such an issue and the comparisons with general RSF and 2D compressible flows will be presented in another communication [24]. As for the dynamics of VGT, checking each terms, coarse-grained/filtered [26] or not [19], for consistent scenario and extra information of our LSF may also shed light on the modeling issue (for large-eddy simulations, say). And we conclude with the expetation that the theoretical framework and computation tool of LSF established here can also be soon applied to fundamental physical problems such as (linear/kinematic) dynamo, aeroacoustics and combustion, among other nonlinear mathematics, the comparisons of which with those associated to RSF and NSF should be very illuminating, while the extensions to quantum flows mentioned in the beginning of the introductory discussions may be more nontrivial due to the fact that the phase defects can make the space multiply connected.